\documentclass[aps,prl,preprint,tightenlines,superscriptaddress,showpacs,byrevtex]{revtex4}
\usepackage{graphicx} 
\usepackage{dcolumn}  

\graphicspath{{ps}}

\begin{document}


\preprint{\vbox{ \hbox{   }
                 \hbox{BELLE-CONF-0404}
                 \hbox{ICHEP04 11-0649}
}}

\title{ \quad\\[0.5cm]  Measurements of Branching Fractions and
  Polarization in $B\to K^*\rho$ Decays}


\affiliation{Aomori University, Aomori}
\affiliation{Budker Institute of Nuclear Physics, Novosibirsk}
\affiliation{Chiba University, Chiba}
\affiliation{Chonnam National University, Kwangju}
\affiliation{Chuo University, Tokyo}
\affiliation{University of Cincinnati, Cincinnati, Ohio 45221}
\affiliation{University of Frankfurt, Frankfurt}
\affiliation{Gyeongsang National University, Chinju}
\affiliation{University of Hawaii, Honolulu, Hawaii 96822}
\affiliation{High Energy Accelerator Research Organization (KEK), Tsukuba}
\affiliation{Hiroshima Institute of Technology, Hiroshima}
\affiliation{Institute of High Energy Physics, Chinese Academy of Sciences, Beijing}
\affiliation{Institute of High Energy Physics, Vienna}
\affiliation{Institute for Theoretical and Experimental Physics, Moscow}
\affiliation{J. Stefan Institute, Ljubljana}
\affiliation{Kanagawa University, Yokohama}
\affiliation{Korea University, Seoul}
\affiliation{Kyoto University, Kyoto}
\affiliation{Kyungpook National University, Taegu}
\affiliation{Swiss Federal Institute of Technology of Lausanne, EPFL, Lausanne}
\affiliation{University of Ljubljana, Ljubljana}
\affiliation{University of Maribor, Maribor}
\affiliation{University of Melbourne, Victoria}
\affiliation{Nagoya University, Nagoya}
\affiliation{Nara Women's University, Nara}
\affiliation{National Central University, Chung-li}
\affiliation{National Kaohsiung Normal University, Kaohsiung}
\affiliation{National United University, Miao Li}
\affiliation{Department of Physics, National Taiwan University, Taipei}
\affiliation{H. Niewodniczanski Institute of Nuclear Physics, Krakow}
\affiliation{Nihon Dental College, Niigata}
\affiliation{Niigata University, Niigata}
\affiliation{Osaka City University, Osaka}
\affiliation{Osaka University, Osaka}
\affiliation{Panjab University, Chandigarh}
\affiliation{Peking University, Beijing}
\affiliation{Princeton University, Princeton, New Jersey 08545}
\affiliation{RIKEN BNL Research Center, Upton, New York 11973}
\affiliation{Saga University, Saga}
\affiliation{University of Science and Technology of China, Hefei}
\affiliation{Seoul National University, Seoul}
\affiliation{Sungkyunkwan University, Suwon}
\affiliation{University of Sydney, Sydney NSW}
\affiliation{Tata Institute of Fundamental Research, Bombay}
\affiliation{Toho University, Funabashi}
\affiliation{Tohoku Gakuin University, Tagajo}
\affiliation{Tohoku University, Sendai}
\affiliation{Department of Physics, University of Tokyo, Tokyo}
\affiliation{Tokyo Institute of Technology, Tokyo}
\affiliation{Tokyo Metropolitan University, Tokyo}
\affiliation{Tokyo University of Agriculture and Technology, Tokyo}
\affiliation{Toyama National College of Maritime Technology, Toyama}
\affiliation{University of Tsukuba, Tsukuba}
\affiliation{Utkal University, Bhubaneswer}
\affiliation{Virginia Polytechnic Institute and State University, Blacksburg, Virginia 24061}
\affiliation{Yonsei University, Seoul}
  \author{K.~Abe}\affiliation{High Energy Accelerator Research Organization (KEK), Tsukuba} 
  \author{K.~Abe}\affiliation{Tohoku Gakuin University, Tagajo} 
  \author{N.~Abe}\affiliation{Tokyo Institute of Technology, Tokyo} 
  \author{I.~Adachi}\affiliation{High Energy Accelerator Research Organization (KEK), Tsukuba} 
  \author{H.~Aihara}\affiliation{Department of Physics, University of Tokyo, Tokyo} 
  \author{M.~Akatsu}\affiliation{Nagoya University, Nagoya} 
  \author{Y.~Asano}\affiliation{University of Tsukuba, Tsukuba} 
  \author{T.~Aso}\affiliation{Toyama National College of Maritime Technology, Toyama} 
  \author{V.~Aulchenko}\affiliation{Budker Institute of Nuclear Physics, Novosibirsk} 
  \author{T.~Aushev}\affiliation{Institute for Theoretical and Experimental Physics, Moscow} 
% ...
\collaboration{The Belle Collaboration}

\noaffiliation

\begin{abstract}
We present results of a study of the charmless vector-vector decays
$B^0 \to K^{*0}\rho^0$ and $B^+\to K^{*0}\rho^+$. The results are
based on a 140 fb$^{-1}$ data sample collected by the Belle  detector
at the KEKB asymmetric  $e^+e^-$ collider.
We obtain the branching fraction 
${\cal B}(B^+\to K^{*0}\rho^+)=(6.6\pm 2.2(\rm stat.) \pm 0.8 (\rm
syst.))\times 10^{-6}$,
and set upper limits on the branching fractions
${\cal B}(B^0 \to K^{*0}\rho^0)<2.6\times 10^{-6}$ and
${\cal B}(B^0 \to f_0(980)K^{*0})<5.2 \times 10^{-6}$.
We also perform a helicity analysis of the $\rho$ and $K^*$ vector
mesons in the decay $B^+\to K^{*0}\rho^+$, 
and obtain the longitudinal polarization fraction
$R_0(B^+\to K^{*0}\rho^+)=0.50\pm 0.19(\rm stat.)^{+0.05}_{-0.07}(\rm
syst.)$.

\end{abstract}


\maketitle

\tighten

{\renewcommand{\thefootnote}{\fnsymbol{footnote}}}
\setcounter{footnote}{0}
In addition to rate asymmetries, $B\to VV$ decays provide
opportunities to search for direct $CP$ and/or $T$
violation~\cite{ref:a.datta} and new physics~\cite{ref:NP}
through angular correlations between the vector meson decay final
states. 
These decays produce final states where three helicity states are
possible, whose amplitudes are called $A_0$, $A_+$ and $A_-$ in the helicity basis.
The Standard Model (SM) with factorization predicts $R_0\gg
R_T$~\cite{ref:vv_polar_kagan, ref:Grossman},
where $R_0=|A_0|^2/(|A_0|^2 +|A_+|^2 + |A_-|^2)$
($R_T=(|A_+|^2+|A_-|^2)/(|A_0|^2 +|A_+|^2 + |A_-|^2)$) is the
longitudinal (transverse) polarization fraction. In
the tree-dominated $B^+\to\rho^+\rho^0$, this prediction is
confirmed~\cite{ref:rhoprho0, ref:Babar_vv}.
In contrast, for the $B\to\phi K^*$ decay, which is a pure $b\to s$
penguin transition, Belle~\cite{ref:phiK} and
Babar~\cite{ref:Babar_vv} find $R_0\approx R_T$, which is in
disagreement with SM predictions. 
It is thus important to obtain
polarization measurements in other VV modes, such as $B \to K^* \rho$
and, in particular, in the pure penguin $b\to s \bar d d$ decay, $B^+\to
K^{*0}\rho^+$.

In this paper, we present the results of a study of $B^0\to
K^{*0}\rho^0$  and $B^+\to K^{*0}\rho^+$ decays with a  140 fb$^{-1}$
data sample containing $152\times 10^{6}$ $B$ meson pairs collected
with the Belle detector at the KEKB asymmetric-energy $e^+e^-$
collider~\cite{KEKB} operating at the $\Upsilon(4S)$ resonance
($\sqrt{s} = 10.58$~GeV). The production rates for $B^+B^-$ and
$B^0\overline{B}{}^0$ pairs are assumed to be equal.

The Belle detector is a large solid-angle magnetic
spectrometer that
consists of a three-layer silicon vertex detector (SVD),
a 50-layer central drift chamber (CDC), an array of
aerogel threshold \v{C}erenkov counters (ACC), 
a barrel-like arrangement of time-of-flight
scintillation counters (TOF), and an electromagnetic calorimeter
comprised of CsI(Tl) crystals (ECL) located inside 
a super-conducting solenoid coil that provides a 1.5~T
magnetic field.  An iron flux-return located outside of
the coil is instrumented to detect $K_L^0$ mesons and to identify
muons (KLM).  The detector is described in detail
elsewhere~\cite{Belle}.

We select $B \to K^* \rho $ candidate events by combining four charged
tracks (three pions plus one kaon) or
three charged tracks (two pions plus one kaon) and one neutral pion.
Each charged track is required to have a transverse momentum $p_T>0.1$
GeV$/c$ and to have an origin within $0.2~{\rm cm}$ in
the radial direction and $5~\rm{cm}$ along the beam direction of the
interaction point (IP).

Particle identification likelihoods for the pion and kaon particle
hypotheses are calculated by combining information from the TOF and ACC
systems with $dE/dx$ measurements in the CDC. To identify kaons, we
require the kaon likelihood ratio, $L_K/(L_K+L_\pi)$, to be greater
than 0.6. To identify pions, we require $L_K/(L_K+L_\pi)$ to be less
than 0.4. In addition, charged tracks are rejected if they are
consistent with the electron hypothesis.

Candidate $\pi^0$ mesons are reconstructed from pairs of photons that
have an invariant mass in the range $0.1178 - 0.1502$
GeV/$c^2$,
corresponding to a window of $\pm 3\sigma$ around the nominal $\pi^0$
mass, where the photons are assumed to originate from the IP.
The energy of each photon in the laboratory frame is required to be
greater than 50 MeV for the ECL barrel region ($32^\circ< \theta
<129^\circ$) and 100 MeV for the ECL endcap regions
($17^\circ<\theta<32^\circ$ or $129^\circ<\theta<150^\circ$), where
$\theta$ denotes the polar angle of the photon with respect to the
beam line. The $\pi^0$ candidates are kinematically constrained to the
nominal $\pi^0$ mass. In order to reduce the combinatorial background,
we only accept $\pi^0$ candidates with momenta $p_{\pi^0}>0.40$
GeV$/c$ in the $e^+e^-$ center-of-mass system (CMS).

Candidate $\rho$ mesons are reconstructed via their 
$\rho^0 \to \pi^+\pi^-$ and $\rho^+ \to \pi^+ \pi^0$ decays.
For both the charged and neutral modes, we require 
$0.62{~\rm GeV}/c^2<M(\pi\pi)<0.92{~\rm GeV}/c^2$.
We select $K^{*0}\to K^+\pi^-$ decay candidates with invariant masses
in the range $0.84{~\rm GeV}/c^2<M(K^+\pi^-)<0.94{~\rm GeV}/c^2$ for
$K^{*0}\rho^0$ and $0.83{~\rm GeV}/c^2<M(K^+\pi^-)<0.97{~\rm GeV}/c^2$ for
$K^{*0}\rho^+$.


To isolate the signal, we form the beam-constrained mass 
$M_{\rm bc}\equiv\sqrt{E_{\rm beam}^2-p_B^2}$, and the energy difference
$\Delta E\equiv E_B-E_{\rm beam}$, where $E_{\rm beam}$ is the CMS
beam energy, and $p_B$ and $E_B$ are the CMS momentum and energy,
respectively, of the $B$ candidate. 
We accept events in the region defined by $M_{\rm bc}>5.2 {~\rm GeV}/c^2$ and
$-0.4{~\rm GeV}<\Delta E<0.4{~\rm GeV}$. Within this accepted range we
further define mode-dependent signal regions.
For $B^0\to K^{*0} \rho^0$, the $\Delta E$
signal region is $-0.04{~\rm GeV}<\Delta E<0.04{~\rm GeV}$.
For $B^+\to K^{*0} \rho^+$, the $\Delta E$ distribution has a tail on 
the lower side caused by incomplete longitudinal containment of
electromagnetic showers in the CsI(Tl) crystals, so the $\Delta E$
signal region is broadened to $-0.10{~\rm GeV}<\Delta E<0.06{~\rm GeV}$.
For both decays, the $M_{bc}$ signal region is $5.27{~\rm
GeV}/c^2<M_{\rm bc}<5.29{~\rm GeV}/c^2$. These requirements correspond to approximately $\pm 3 \sigma$
for both quantities.

The continuum process $e^+e^- \to q \bar{q}$ ($q=u, d, s, c$) is the
main source of background and must be strongly suppressed. 
One method of discriminating the signal from background is based on
the event topology, which tends to be isotropic for $B\bar B$ events
and jet-like for $q\bar q$ events.  Another discriminating
characteristic is $\theta_B$, the CMS polar angle of the $B$ flight
direction.
$B$ mesons are produced with a $1-\cos^2\theta_B$ distribution
while continuum background events tend to be uniform in $\cos\theta_B$.
For $B^0\to K^{*0} \rho^0$,
we require $|\cos\theta_{\rm thr}|<0.8$, where $\theta_{\rm thr}$
is the angle between the thrust axis of the candidate tracks and that
of the remaining tracks in the event.  This distribution is flat for
signal events and peaked at $\cos\theta_{\rm thr}=\pm 1$ for continuum
background.

We use Monte Carlo (MC) simulated signal and continuum
events to form a
Fisher discriminant based on modified Fox-Wolfram moments~\cite{ref:fox}
that are verified to be uncorrelated with $M_{\rm bc}$, $\Delta E$ and
variables considered later in the analysis.
Probability density functions (PDFs) derived from the
Fisher discriminant and the $\cos\theta_B$ distributions are
multiplied to form likelihood functions for signal (${\cal L}_{s}$)
and continuum (${\cal L}_{q\bar q}$);
these are combined into a likelihood ratio 
${\cal R}_s={\cal L}_{s}/({\cal L}_{s}+{\cal L}_{q\bar q})$.
Additional discrimination is provided by the $b$-flavor
tagging parameter $r$, which ranges from 0 to 1 and is
a measure of the likelihood that the $b$ flavor 
of the accompanying $B$ meson is correctly assigned
by the Belle flavor-tagging algorithm~\cite{ref:sin2phi1}. 
Events with high values of $r$ are well-tagged and are less
likely to originate from continuum production.
We define a multi-dimensional likelihood ratio ${\mathcal M}= {\cal 
L}_{s}^{MDLR}/({\cal L}_{s}^{MDLR}+ {\cal 
L}_{q\bar q}^{MDLR})$, where ${\cal L}_{s}^{MDLR}$
denotes the likelihood determined by the $r$-${\cal R}_s$
distribution for signal and ${\cal L}_{q\bar q}^{MDLR}$
is that for the continuum background. 
We determine ${{\mathcal M}}$ cut by optimizing the figure of merit,
$S/\sqrt{(S+B)}$, where $B$ is the number of background events and $S$
is the number of signal events.
We require ${\mathcal M}>0.85$ for $K^{*0}\rho^0$ and ${\mathcal M}>0.95$ for
$K^{*0} \rho^+$.


To investigate backgrounds from other $B$ decays, we use a sample of
 $B\bar B$ MC events corresponding to an integrated luminosity of 412
 fb$^{-1}$. We find background from $B^+ \to \overline
 D{}^0(K^{*0}\pi^0)\pi^+$  in the $\rho$ sideband region. We apply the
 requirement 
$|M(K\pi^+\pi^0)-M_{D^0}|>0.050$ GeV$/c^2$ to veto those events. 
This requirement does not remove any $B^+\to K^{*0}\rho^+$
events. The $B^0\to K^{*0}\rho^0$ mode includes vetoes on $B^0\to 
D^-(K^-\pi^+\pi^-)\pi^+$, $B^0\to D^{*}(2010)^-(D^0\pi^-)\pi^+$ and
 $B^0\to D^{*}_2(2460)^-(D^0\pi^-)\pi^+$.
The cuts $|M(K\pi\pi)-M_{D^0}|>0.050$ GeV$/c^2$,
 $|M(K\pi\pi)-M_{D^*(2010)}|>0.050$ GeV$/c^2$ and $|M(K\pi\pi)-M_{D_2^*(2460)}|>0.060$
 GeV$/c^2$ eliminate these backgrounds.
  
After all the event selection requirements are applied, there is no
significant peaking background in either $\Delta E$ or $M_{\rm bc}$ for
$B^+\to K^{*0} \rho^+$. 
For $B^0\to K^{*0}\rho^0$ we find some broadly peaking 3-body and 5-body
rare decays
above and below the signal
in the $\Delta E$ distribution. These shapes
are included in the $M_{\rm bc}$-$\Delta E$ fit for this mode. A
further cut, $\Delta E>-0.2{~\rm GeV}$, is also applied to eliminate
most of the remaining 5-body background. 

In the $B^+\to K^{*0} \rho^+$ $M_{\rm bc}$-$\Delta E$ signal region,
we find that the fraction of multiple candidates is 3.5\% for the
$|A_0|^2$ helicity state and 1.6\% for the $|A_\pm|^2$ state. We allow
for multiple candidates in this mode.
For $B^0\to K^{*0} \rho^0$ the wide mass windows result in many
multiple candidates; we choose the entry that has the minimum $\chi^2$
from the vertex constrained fit.

We extract the signal yield by applying an extended unbinned
maximum-likelihood fit to the two-dimensional $M_{\rm bc}$-$\Delta E$
distribution. The fit includes components for signal plus backgrounds
from continuum events and $b\to c$ decays.
The signal PDF is represented by a Gaussian function for $M_{\rm bc}$
and either a double Gaussian (for the $K^{*0}\rho^0$ mode) or a ``Crystal Ball'' line shape 
function~\cite{ref:cb} (for the $K^{*0}\rho^+$ mode) for $\Delta E$.
The shape parameters are determined from fits to
MC. The signal PDF is adjusted to account for small differences
observed between data and MC. 
The $B^0\to K^{*0} \rho^0$ signal PDF is calibrated with $B^0 \to
D{}^0(K^-\pi^+\pi^-) \pi^+$;
the $B^+\to K^{*0}\rho^+$ PDF is calibrated with high-statistics modes
containing $\pi^0$ mesons, i.e., $B^+ \to {\overline D}{}^0(K^+\pi^-\pi^0)\pi^+$.
The continuum PDF is described by a product of a threshold (ARGUS)
function~\cite{ref:ARGUS} for $M_{\rm bc}$ and a first-order
polynomial for $\Delta E$, with shape parameters allowed to vary.
The PDF for $b\to c$ decay is modeled by a smoothed two-dimensional
histogram obtained from a large MC sample. 
In addition, the $B^0\to K^{*0} \rho^0$ fit contains PDFs for the
dominant rare 3-body ($B^0\to K^{*0}\pi$) and 5-body ($B^0\to a_1
K^{*0}$) decays, modeled by smoothed two-dimensional histograms.

\begin{figure}[htbp]
  \begin{center}
\includegraphics[width=12cm,clip]{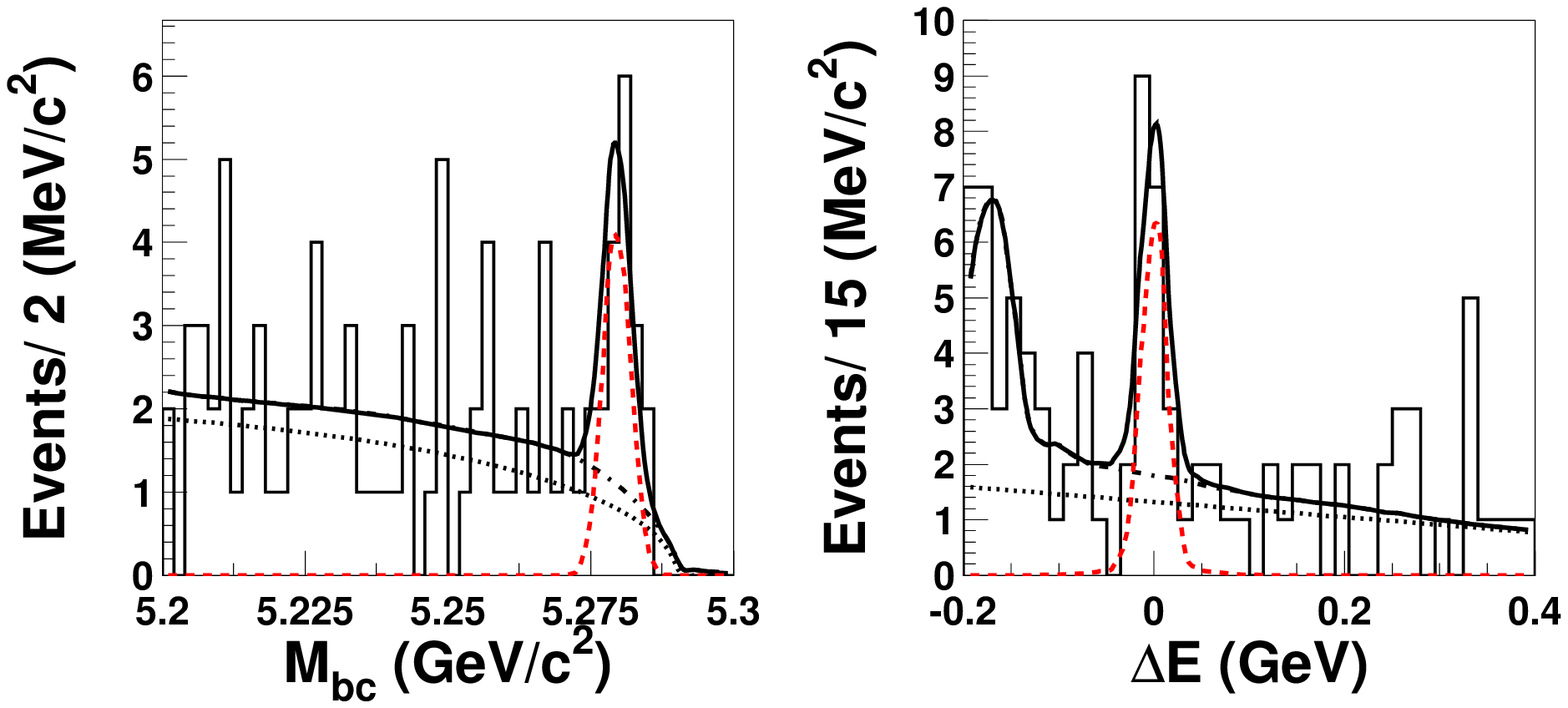} 
\vspace{2mm}\\
    \includegraphics[width=12cm,clip]{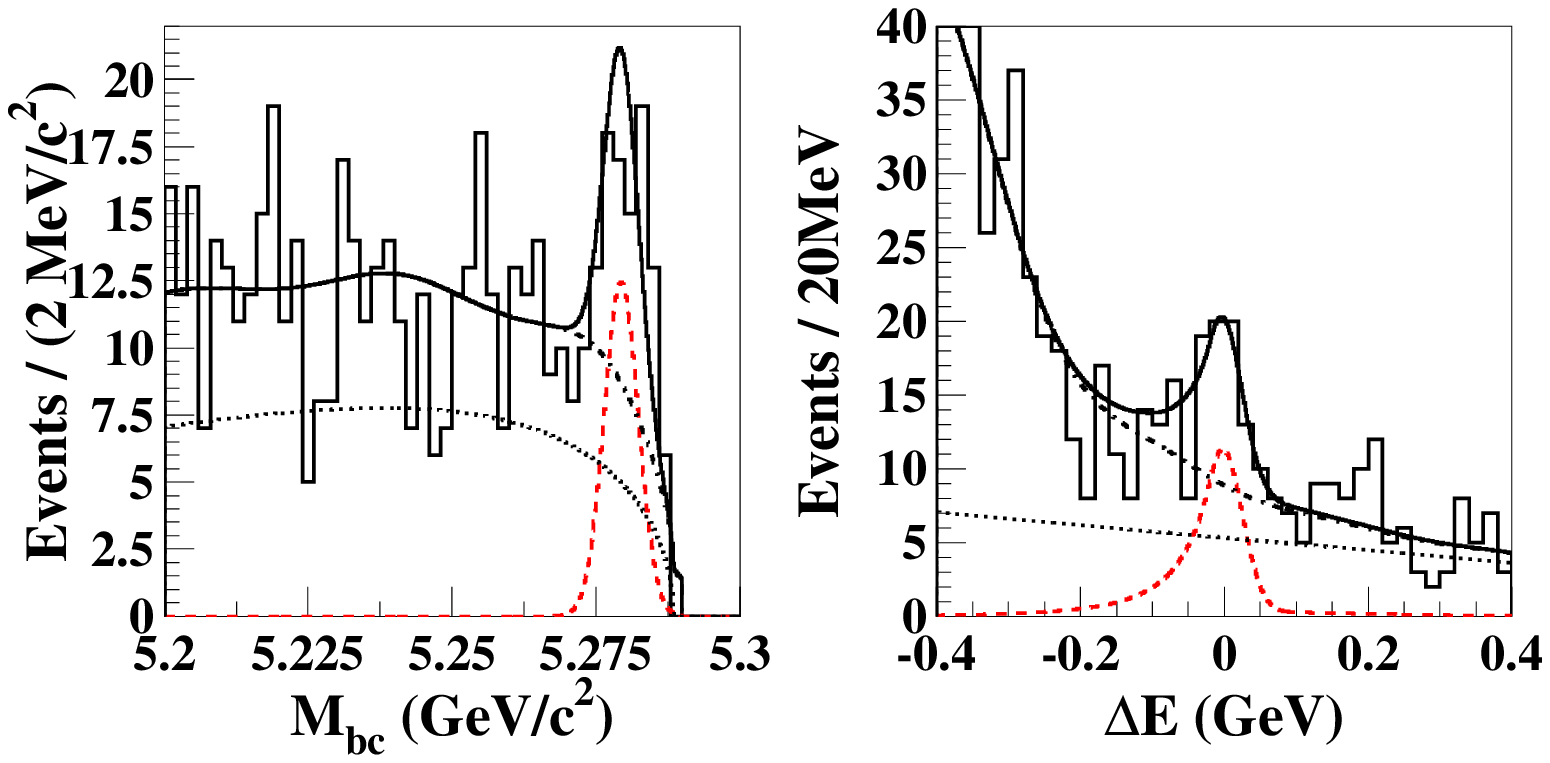}
    \caption{Projections of $M_{\rm bc}$ for events in the $\Delta E$
    signal region (left), and projection of $\Delta E$ in the $M_{\rm
    bc}$ signal region (right). The upper plots are for $B^0\to K^{*0}
    \rho^0$, the lower plots for $B^+\to K^{*0} \rho^+$. The solid curves
    show the results of the fits. The dotted lines represent the
    continuum background. The sum of $b\to c$ and continuum background
    component is shown as dashed lines.}
    \label{fitdata_proj}
  \end{center}
\end{figure}

Figure~\ref{fitdata_proj} shows the fit results. We find $14.1\pm 4.4$
$B^0\to K^+\pi^-\pi^+\pi^-$ events and $56.5\pm 11.6$ $B^+\to
K^+\pi^-\pi^+\pi^0$ events.
The corresponding statistical significance of the signal, defined as
$\sqrt{-2\ln({\cal L}_0/{\cal L}_{\rm max})}$, where ${\cal L}_{\rm
  max}$ is the likelihood value at the best-fit signal  yield and
${\cal L}_{0}$ is the value with the signal yield set to zero, is
5.0$\sigma$ or 6.3$\sigma$.

The $M_{\rm bc}-\Delta E$ fits do not distinguish signal from
non-resonant decays, such as $B\to \rho K\pi$ or $B\to K^{*}\pi\pi$,
which contain the same final state particles.
Therefore, we extract the $K^* \rho$ signal yield by fitting the $M(\pi\pi)$ and
$M(K\pi)$ invariant masses for events in the $M_{\rm bc}$ and $\Delta
E$ signal region.

\begin{figure}[htbp]
\begin{center}
    \includegraphics[width=12cm,clip]{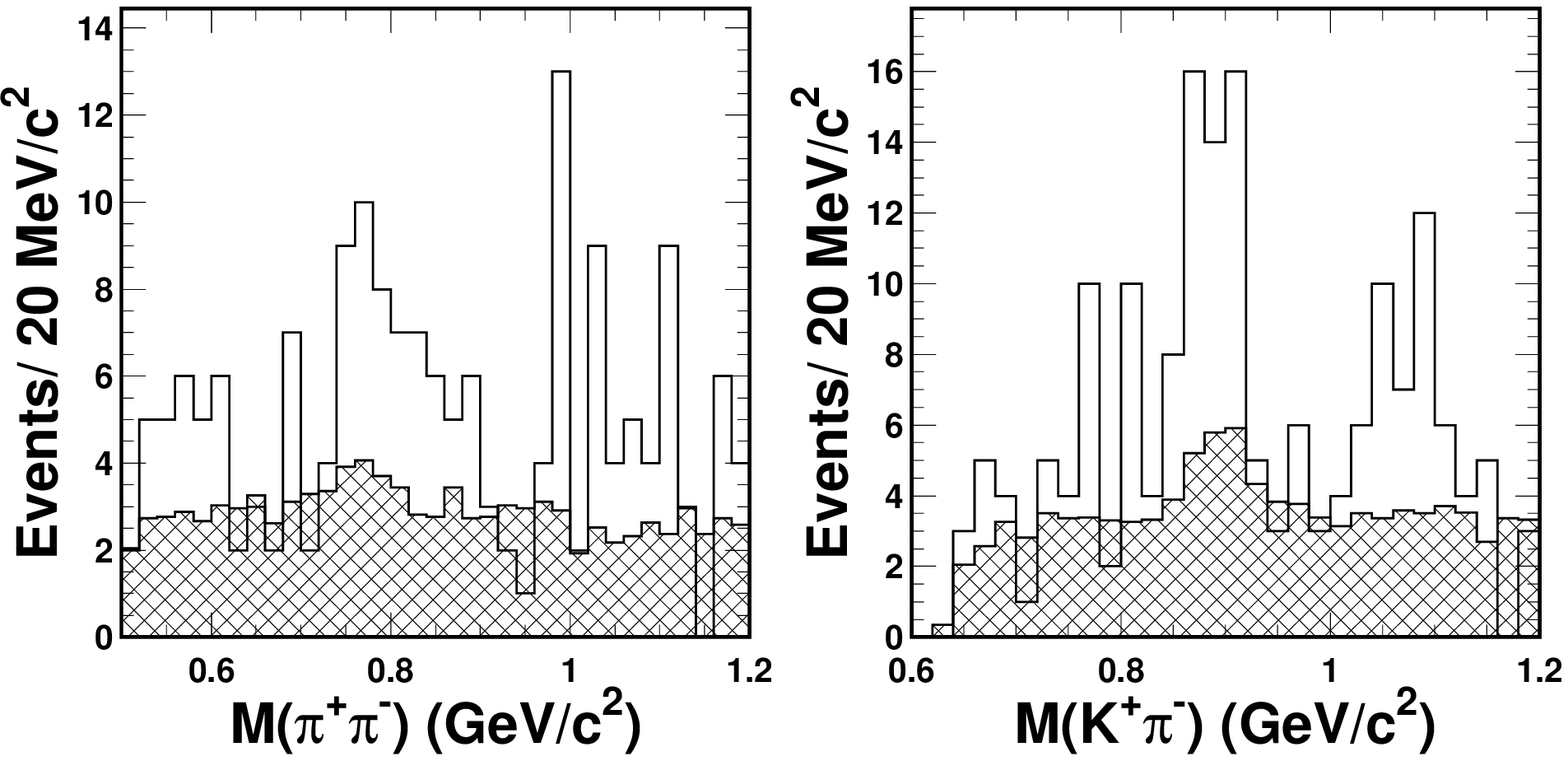} 
      \caption{Distributions of $M(\pi^+\pi^-)$(left) and $M(K^+\pi^-)$
      (right) in the $K^{*0} \rho^0$ signal region (open histogram)
      and $M_{\rm bc}$ sideband (hatched histogram). Peaks for
      $\rho^0$ and $K^{*0}$ can clearly be seen and a sharp peak for
      the $f_0(980)$ is apparent. Significant contributions of
      non-resonant background are also seen. }
    \label{fig:mass_spectrum}
  \end{center}
\end{figure}

\begin{figure}[htbp]
\begin{center}
    \includegraphics[width=12.5cm,clip]{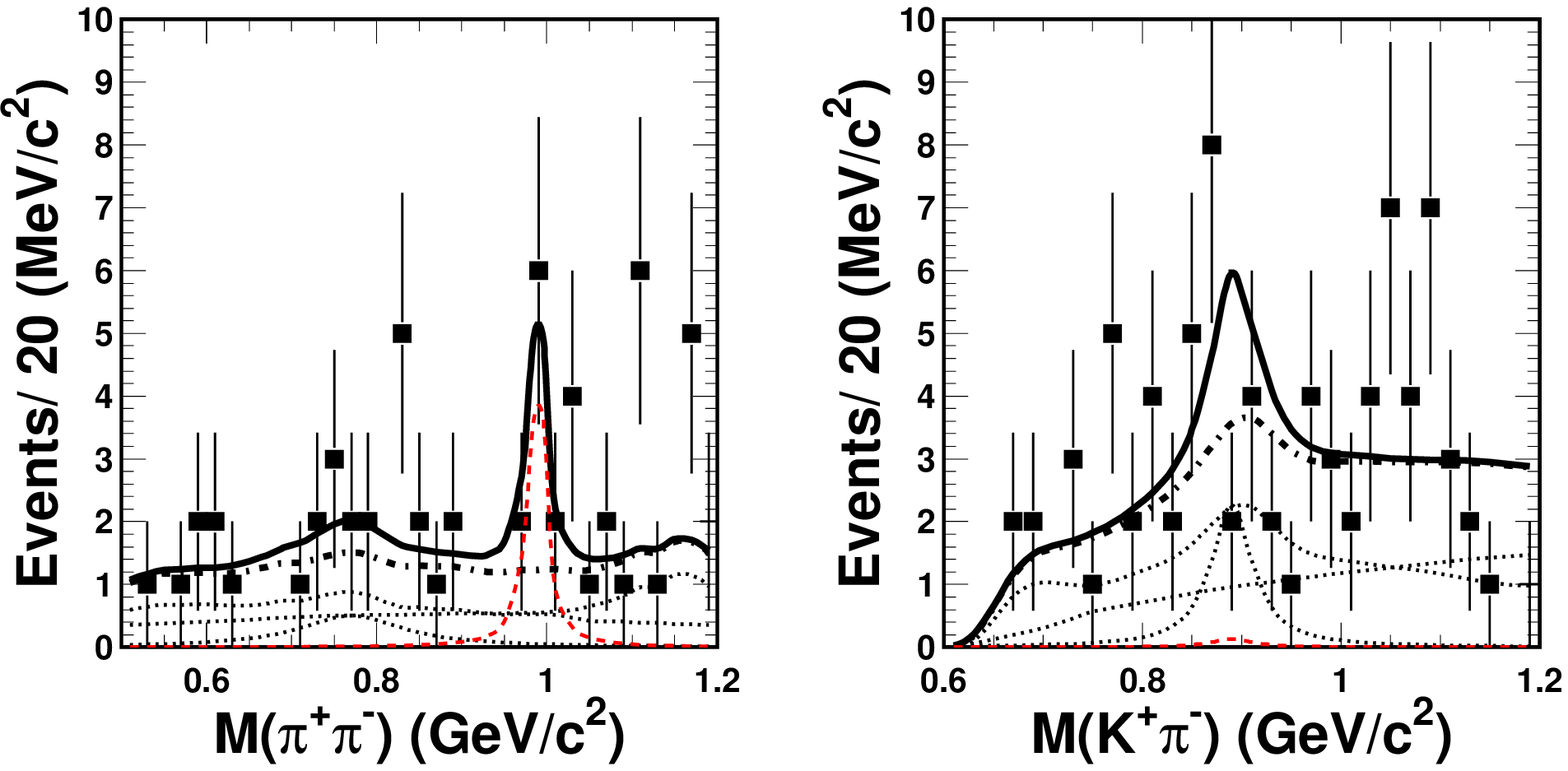} 
    \includegraphics[width=6cm,clip]{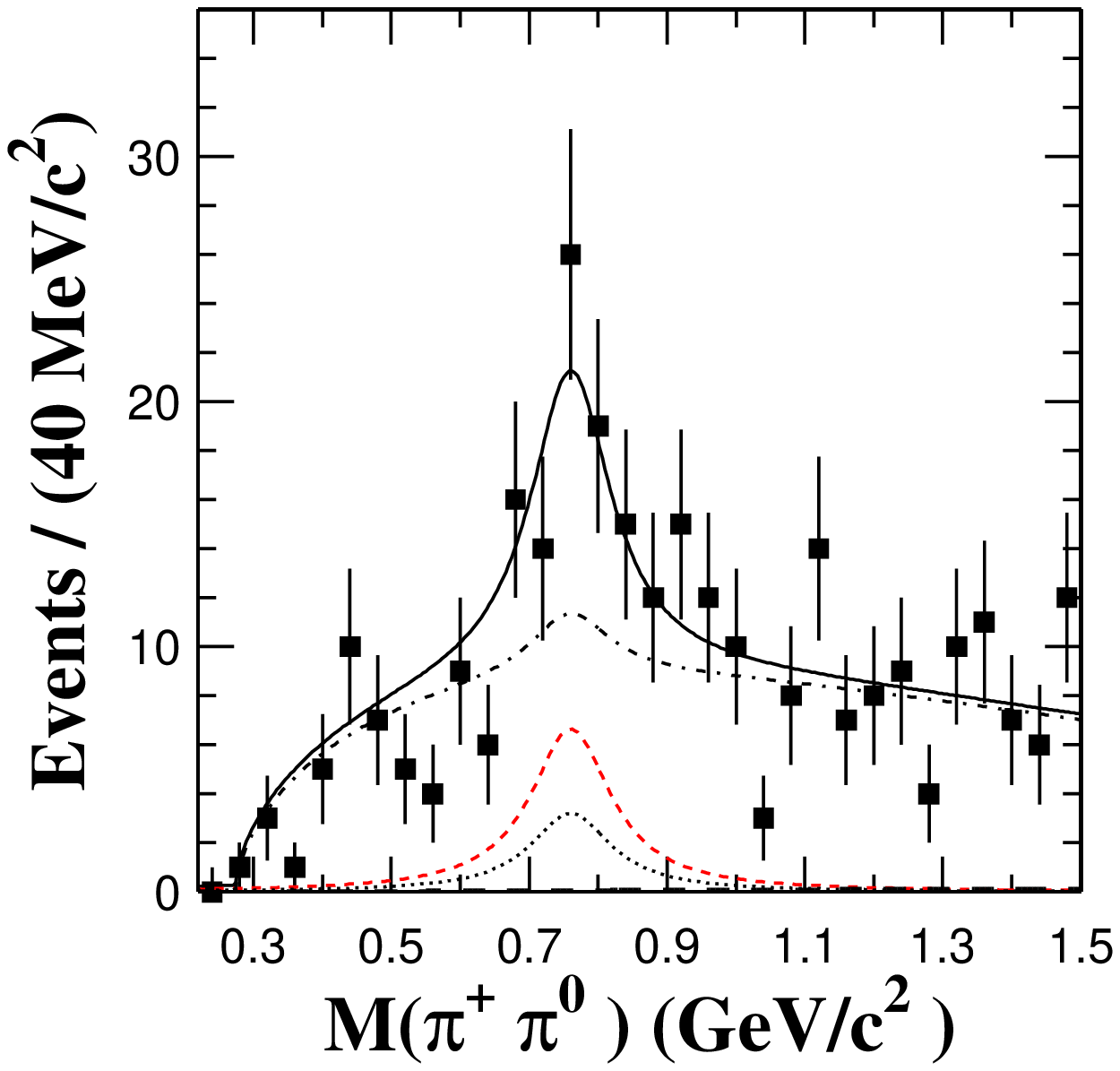}
    \includegraphics[width=6cm,clip]{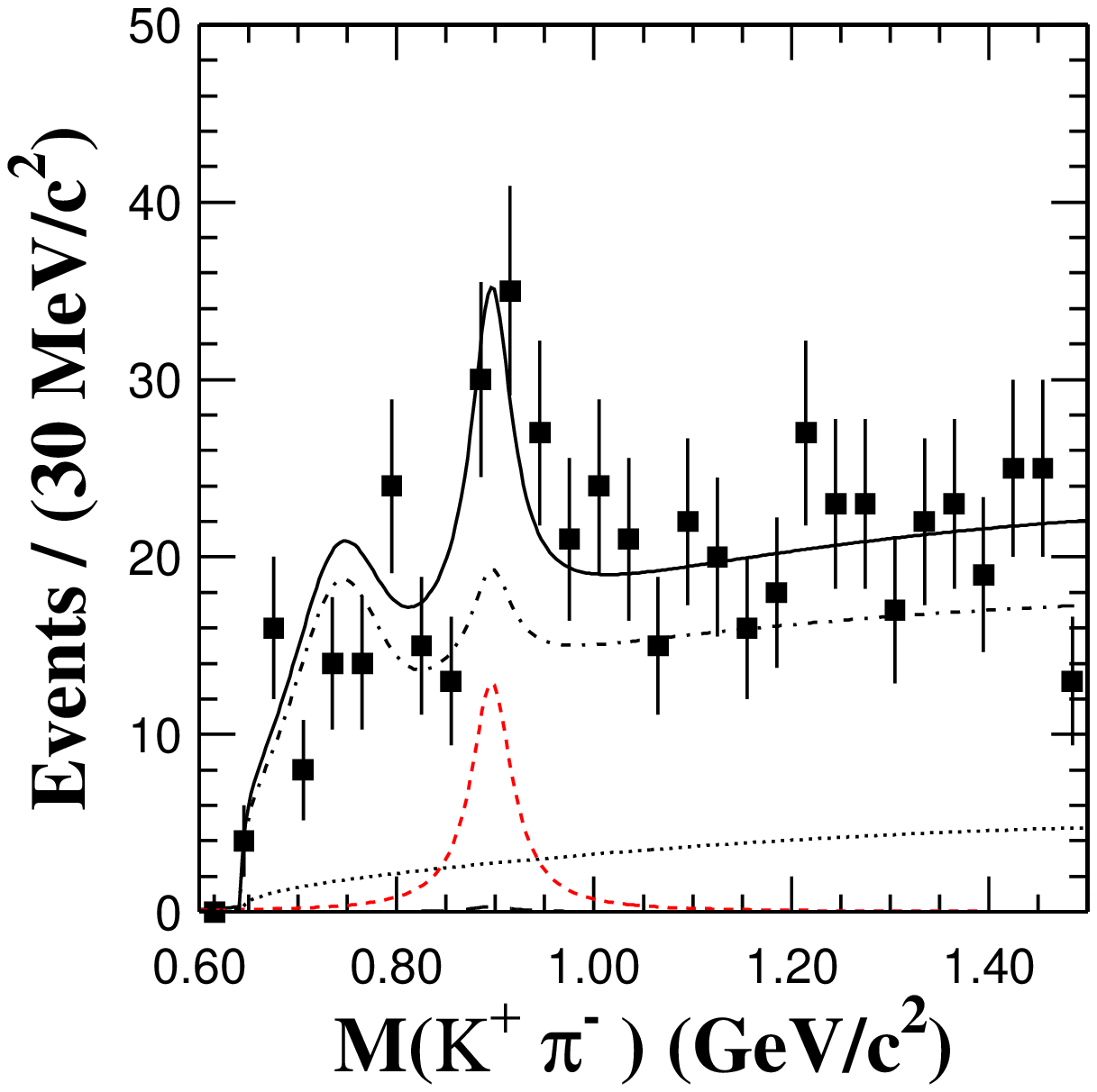}
    \caption{The upper plots are for $B^0\to K^{*0}\rho^0$,
      $M(\pi^+\pi^-)$ (left) in the $K^{*0}$ signal region and
      $M(K^+\pi^-)$ in the $\rho$ signal region (right). 
      Solid curves show fit results. Dotted lines show continuum and
      non-resonant components, red dashed line shows the
      $f_0(980)K^{*0}$ component.
      The lower plots are for $B^+\to K^{*0}\rho^+$,
      $M(\pi^+\pi^0)$ (left) in the $K^{*0}$ signal region
      and $M(K^+\pi^-)$ in the $\rho$ signal region (right).
      The solid curves show the results of the fit.
      The signal (continuum, $\rho K\pi$,  $K^{*0}\pi\pi$ ) component
      is shown as a red dashed (dot-dashed, dotted, dashed) line.}
    \label{fig:fit_mass}
  \end{center}
\end{figure}

For $B^0\to K^{*0} \rho^0$ signal region events, we perform a
two-dimensional maximum likelihood fit in
the range $0.5{~\rm GeV}/c^2<M(K^+\pi^-)<1.2{~\rm GeV}/c^2$ and
$0.5{~\rm GeV}/c^2<M(\pi^+\pi^-)<1.2{~\rm GeV}/c^2$. The distribution
for these events is shown in Fig.~\ref{fig:mass_spectrum}. From
this distribution it is clear there are significant contributions from
non-resonant backgrounds.
The signal PDF is modeled by the product of Breit-Wigner functions
for both the $\rho^0$ and $K^{*0}$ resonances with means and widths
determined from the MC simulation. The continuum and $b\to c$ fractions
are each modeled by a smoothed two-dimensional histogram determined
from MC. Three main non-resonant decays are considered in this fit: 
$B^0\to K^{*0}\pi^+\pi^-$, $B^0\to K_1(1400)^+(K^{*0}\pi^+)\pi^-$ and
$B^0\to \rho^0 K^-\pi^+$.  
The $K^{*0}\pi^+\pi^-$ and $\rho^0 K^-\pi^+$ components
are represented as a Breit-Wigner function for the resonance,
multiplied by a second order polynomial fit to the non-resonant phase
space, determined from MC. The $K_1(1400)^+\pi^-$ component is modeled
by a smoothed histogram.
From the data mass distribution it is apparent that an
additional component to model the $B^0\to f_0(980)K^{*0}$ decay is needed.
This is represented by the product of two Breit-Wigner functions for
the $K^{*0}$ and $f_0(980)$ resonance. The $K^{*0}$ mean and width
are determined from MC. The $f_0(980)$ mean and width are allowed to
float within the range of values reported in the PDG~\cite{ref:PDG}.
The continuum and $b\to c$ backgrounds are fixed at the levels
determined from the $M_{\rm bc}$ - $\Delta E$ fit; the other components
are allowed to float. The projections of the fit are shown in  
Fig.~\ref{fig:fit_mass}. Table~\ref{tab:kst0rho0_mass} lists the
results of the fit. The statistical significance of the
$f_0(980)K^{*0}$ signal is 2.8$\sigma$.

For $B^+\to K^{*0} \rho^+$, we fit $M(\pi\pi)$ in the $K^*$
signal region and $M(K\pi)$ in the $\rho$ signal region simultaneously.
Figure~\ref{fig:fit_mass} (bottom) shows the fit result.
The solid curve shows the result of a binned maximum-likelihood fit
with four components: signal, background from continuum and $B\bar B$, and
non-resonant $\rho K\pi$ and $K^{*0}\pi\pi$. The signal $\rho$ and $K^{*0}$
components are represented by Breit-Wigner functions with masses and
widths determined from MC simulation.
The background from continuum and $B\bar B$ is described by a
threshold function plus a Breit-Wigner function for $M(\pi^+\pi^0)$,
a threshold function plus a Gaussian and a Breit-Wigner function for
$M(K^+\pi^-)$, where a resonant component is included to account for
resonance production in the continuum.
The shape parameters are determined from sideband data.
The non-resonant $\rho K\pi$ and $K^*\pi\pi$ components are
represented by threshold functions with parameters determined from
a MC simulation with the final state particles distributed uniformly
over phase space.
In the fit, the yield for each component in the $\rho$ or $K^{*0}$
signal region is required to be the same. All normalizations are
allowed to float, except for the background from continuum and $B\bar
B$, which is fixed at the $M_{\rm bc}$-$\Delta E$ fit results.
Table~\ref{tab:kst0rhop_mass} lists the yields in the $\rho^+$ and
$K^{*0}$ mass window.
The statistical significance of the $B^+\to K^{*0}\rho^+$ signal is
$3.2\sigma$.

\begin{table}[htbp]
  \begin{center}
    \caption{Results of fit to invariant mass for $B^0\to
    K^{*0}\rho^0$. The yields are the number of events over the whole
    $M(\pi\pi)$ - $M(K\pi)$ region.}
    \begin{tabular}{ccccccc} \\ \hline \hline
      \hspace*{0.5cm } $K^{*0}\rho^0$ &
      \hspace*{0.5cm } $f_0(980)K^{*0}$ &
      \hspace*{0.5cm } $\rho K\pi$ &
      \hspace*{0.7cm } $K^{*0}\pi\pi$ &
      \hspace*{0.7cm } $K_1 \pi$ &
      \hspace*{0.2cm }$|A_0|^2$ effic.(\%) &
      \hspace*{0.2cm } $|A_\pm|^2$ effic.(\%) \\
      \hline
      $0\pm 5.2$ & $10.2^{+5.3}_{-4.4}$ & $30.5^{+10.7}_{-9.8}$ &
      $22.4^{+13.0}_{-12.2}$ & $9.4^{+9.4}_{-8.6}$ &1.9 & 3.4\\
      \hline \hline
    \end{tabular}
    \label{tab:kst0rho0_mass}
  \end{center}
\end{table}

\begin{table}[htbp]
\begin{center}
\caption{Results of fit to invariant mass for $B^+ \to K^{*0} \rho^+$,
  together with the MC-determined efficiencies for $|A_0|^2$ and
  $|A_\pm|^2$ states. The yields are the number of events in the $K^*$
  and $\rho$ mass window.}
\begin{tabular}{ccccc} \\ \hline \hline
\hspace*{0.5cm } $K^{*0}\rho^+$ \hspace*{0.5cm } &
\hspace*{0.5cm } $\rho K\pi$    \hspace*{0.5cm } &
\hspace*{0.5cm } $K^* \pi\pi$   \hspace*{0.5cm } \hspace*{0.5cm}&
\hspace*{0.5cm }$|A_0|^2$ effic.(\%) &
\hspace*{0.5cm } $|A_\pm|^2$ effic.(\%)\\
\hline
$26.6\pm 8.7$  &  $12.8\pm3.3$  & $0.6\pm 3.7$ & 2.0   & 3.3 \\ 
\hline \hline
\end{tabular}
\label{tab:kst0rhop_mass}
\end{center}
\end{table}

\begin{figure}[htbp]
  \begin{center}
    \includegraphics[width=12cm,clip]{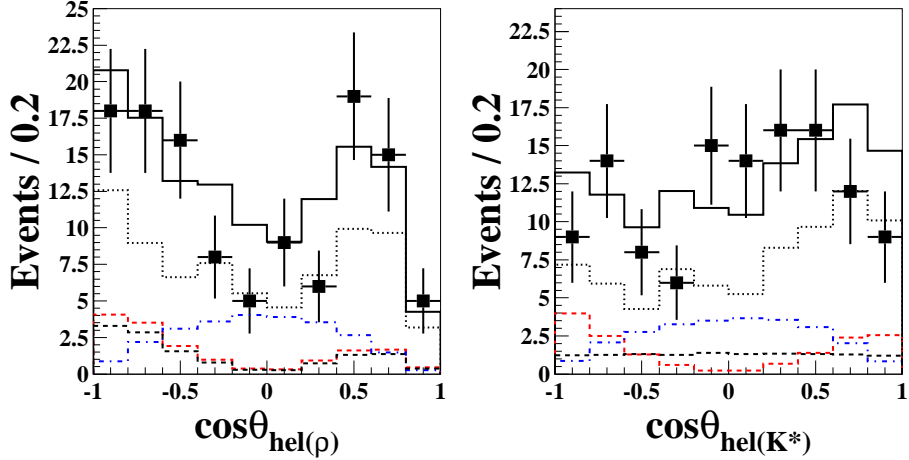}
    \caption{Projections to $\cos \theta_{\rm hel(\rho)}$ (left) and $\cos
      \theta_{\rm hel (K^*)}$ (right).
      The solid histograms show the results of the two-dimensional fit.
      The red dashed (blue dot-dashed) histograms are the $A_0$
      ($A_\pm$) component of the fit; 
      The dotted histograms are backgrounds from continuum and $B\bar B$.
      The black dashed histograms are for non-resonant $\rho K\pi$.
      The low event yield near $\cos\theta_{{\rm
      hel}(\rho)}=1$ is due to the $p_{\pi^0}>0.4{\,\rm GeV}/c$
      requirement.}
    \label{fig:helfit}
  \end{center}
\end{figure}

We use the $\rho^+ \to\pi^+\pi^0$ and $K^{*0}\to K^+\pi^-$
helicity-angle ($\theta_{\rm {hel}}$) 
distributions to determine the relative strengths of $|A_{0}|^2$ and
$|A_{\pm}|^2$. Here $\theta_{\rm {hel}}$ is the angle between
an axis anti-parallel to the $B$ flight direction and the
$\pi^+$ ($K^+$) flight direction in the $\rho^+$ ($K^*$) rest frame. 
For the longitudinal polarization case, the distribution  is
proportional to $\cos ^2 \theta_{{\rm hel}(\rho)} \cos^2\theta_{{\rm
    hel}(K^*)}$ and, for the transverse polarization case, is proportional to
$\sin^2\theta_{{\rm hel}(\rho)} \sin^2\theta_{{\rm hel}(K^*)}$, where
$\theta_{{\rm hel}(\rho)}$ ($\theta_{{\rm hel}(K^*)}$) is the
helicity angle for $\rho$ ($K^*$).
Figure~\ref{fig:helfit} shows the distributions of the cosine of the
helicity angle for $\rho$ and for $K^*$ for events in
the $M_{\rm bc}$ and $\Delta E$ signal region.
We perform a two-dimensional unbinned maximum likelihood fit to the
$\rho$ and $K^*$ helicity distributions.
The fit includes components for signal, backgrounds from continuum and
$B\bar B$ and non-resonant $\rho K\pi$. 
The $K^*\pi\pi$ component obtained from the invariant mass fit is
small and is not included in the helicity fit.
PDFs for signal $|A_0|^2$, $|A_\pm|^2$ helicity states are determined
from the MC simulation.
The PDF for background from continuum and $B\bar B$ is obtained from
sideband data.
The $\rho K\pi$ PDF is determined by fitting the helicity distribution
for events in the  region $1.0~{\rm GeV}/c^2<M(K\pi)<1.5$
GeV/$c^{-2}$; it is
consistent with a $\cos^2\theta_{\rm hel}$-like $\cos \theta(\rho)$
helicity and a flat $\cos \theta(K\pi)$ distribution. 
The fit result, shown as the solid histogram, gives $R_0=0.50\pm
0.19$; $3.2\sigma$ away from 100\% longitudinal polarization.

We obtain the longitudinal polarization fraction,
$$R_0(B^+\to K^{*0}\rho^+)= 0.50\pm 0.19(\rm stat.)^{+0.05}_{-0.07}(\rm
syst.),$$
where the systematic error is associated with the fitting procedure
determined by shifting each parameter by $\pm 1\sigma$, changing the PDF,
and taking the quadratic sum of the resulting changes in $R_0$ as the
systematic error.

We use the mass fit result and MC-determined
efficiencies weighted by the measured polarization components to
calculate the branching fraction for $B^+ \to K^{*0}\rho^+$;
for the calculation of the upper limit of $B^0 \to K^{*0}\rho^0$,
100\% longitudinal polarization is assumed.

We consider systematic errors in the branching fraction of the decay
$B \to K^*\rho$ that are caused by uncertainties in the efficiencies
of track finding, particle identification, $\pi^0$ reconstruction,
continuum suppression and fitting.
We assign a 1.1\%/track error for the uncertainty in the tracking
efficiency. This uncertainty is obtained from a study of partially
reconstructed $D^*$ decays.
We also assign a $0.6$\%/track error for the particle identification
efficiency that is based on a study of kinematically selected $D^{*+}
\to D^0\pi^+$, $D^0\to K^-\pi^+$ decay.
A 4.0\% systematic error for the uncertainty in the $\pi^0$ detection
efficiency is determined from data-MC comparisons of
$\eta\to\pi^0\pi^0\pi^0$ with $\eta \to\pi^+\pi^-\pi^0$ and $\eta \to
\gamma\gamma$.
A 3.4\% ($K^{*0}\rho^0$)/ 3.8\% ($K^{*0}\rho^+$) systematic error for
continuum suppression is estimated from studying $B^+ \to
D^-\pi^+$, $D^-\to K^-\pi^+\pi^-$ and $B^+ \to \overline{D}{}^0\pi^+$,
$\overline{D}{}^0\to K^+\pi^-\pi^0$.
A 3.8\% systematic error associated with the $K^{*0}\rho^+$ fit is obtained by
shifting the parameters by $\pm 1\sigma$ and changing the PDF.
A fit systematic error of 16.1\% for $K^{*0}\rho^0$ and 7.8\% for
$f_0(980)K^{*0}$ is found by varying the fit parameters by $\pm
1\sigma$.
A 0.5\% error for the uncertainty in the number of $B\bar{B}$ events
in the data sample is also included.
For $B^+\to K^{*0}\rho^+$, we also include a $11.9\%$ error due to
the uncertainty in the fraction of longitudinal polarization. 
The quadratic sum of all of these errors is take as the total
systematic error.
We obtain the branching fraction
$${\cal B}(B^+\to K^{*0}\rho^+) =(6.6 \pm 2.2(\rm stat.)\pm 0.8(\rm
syst.)) \times 10^{-6},$$
and set 90\% confidence level (C.L.) upper limits on
$B^0\to K^{*0}\rho^0$,
$${\cal B}(B^0\to K^{*0}\rho^0)<2.6 \times 10^{-6},$$
and $B^0\to f_0(980)K^{*0}$,
$${\cal B}(B^0\to f_0(980)K^{*0})<5.2 \times 10^{-6}.$$

In summary, we measure the branching fraction for $B^+\to
K^{*0}\rho^+$, set a 90\% CL. upper limit on $B^0\to K^{*0}\rho^0$ and
$B^0\to f_0(980)K^{*0}$. 
A helicity analysis is performed for $B^+\to K^{*0}\rho^+$.
We find a substantial transversely polarized fraction.
The results are consistent with some recent
calculations~\cite{ref:vv_polar_kagan, ref:vv_polar_other}.
All results reported here are preliminary.

We thank the KEKB group for the excellent operation of the
accelerator, the KEK Cryogenics group for the efficient
operation of the solenoid, and the KEK computer group and
the National Institute of Informatics for valuable computing
and Super-SINET network support. We acknowledge support from
the Ministry of Education, Culture, Sports, Science, and
Technology of Japan and the Japan Society for the Promotion
of Science; the Australian Research Council and the
Australian Department of Education, Science and Training;
the National Science Foundation of China under contract
No.~10175071; the Department of Science and Technology of
India; the BK21 program of the Ministry of Education of
Korea and the CHEP SRC program of the Korea Science and
Engineering Foundation; the Polish State Committee for
Scientific Research under contract No.~2P03B 01324; the
Ministry of Science and Technology of the Russian
Federation; the Ministry of Education, Science and Sport of
the Republic of Slovenia; the National Science Council and
the Ministry of Education of Taiwan; and the U.S.\
Department of Energy.


\end{document}